\documentclass[jocse]{jocseart}

\usepackage{caption}
\usepackage{subcaption}
\usepackage{booktabs} % For formal tables
\usepackage{fancyhdr}
\usepackage[inline]{enumitem}
\begin{document}

\title{Cloud to Ground Secured Computing: User Experiences on the Transition from Cloud-Based to Locally-Sited Hardware}

\author{Carolyn Ellis}
\affiliation{%
  \institution{Purdue University}
  \city{West Lafayette, IN}
  \country{USA}}
\email{carolynellis@purdue.edu}

\author{Matthew Route}
\affiliation{%
    \institution{Purdue University}
    \city{West Lafayette, IN}
    \country{USA}}
\affiliation{%
	\institution{XSEDE Domain Champion for Astrophysics, Aerospace, and Planetary Science}}
\email{mroute@purdue.edu}

\author{Christopher Phillips}
\affiliation{%
    \institution{Purdue University}
    \city{West Lafayette, IN}
    \country{USA}}
\email{phill219@purdue.edu}

\author{Nicholas Smith}
\affiliation{%
    \institution{Purdue University}
    \city{West Lafayette, IN}
    \country{USA}}
\email{smithnp@purdue.edu}

\author{Alex Younts}
\affiliation{%
    \institution{Purdue University}
    \city{West Lafayette, IN}
    \country{USA}}
\email{ayounts@purdue.edu}
%Others?

%\renewcommand{\shortauthors}{Foo, et al.}

\begin{abstract}
The application of high-performance computing processes, tools, and technologies to Controlled Unclassified Information (CUI) creates both opportunities and challenges.  Building on our experiences developing, deploying, and managing the Research Environment for Encumbered Data (REED) hosted by AWS GovCloud, Research Computing at Purdue University has recently deployed Weber\cite{weber}, our locally-sited HPC solution for the storage and analysis of CUI data.  Weber presents our customer base with advances in data access, portability, and usability at a low, stable cost while reducing administrative overhead for our information technology support team.
\end{abstract}



\maketitle

\section{Introduction}

\subsection{Background}
In 2015-2016, Purdue University built it's first environment for Controlled Unclassified Information (CUI) data subject to NIST SP 800-171\cite{nist800} requirements, REED (Research Environment for Encumbered Data). REED was built in AWS GovCloud because of physical space limitations and time restraints to get it built and deployed. AWS GovCloud restricts access to users within the geographical boundaries of the United States of America and can host sensitive and regulated data by complying with International Traffic in Arms Regulations (ITAR), or requiring compliance with NIST SP 800-171. Through it's lifespan, REED hosted a handful of Windows terminal type project spaces, and a single, traditional HPC project with node scheduling. REED served as an early example within the Higher Education space for environments handling CUI with an Educause white paper \cite{Steinwhitepaper} and $\sim$40 consultations with other institutions handling these cybersecurity requirements. After REED had stabilized in terms of management processes and technologies, Purdue Research Computing staff held sessions for feedback to evaluate REED in terms of control implementation, researcher ease of use, operations, and cost perspectives.

\subsection{Feedback Summarized}

The primary detractor during REED's operation was the cost. Another problem which compounded this was that faculty had grown accustom to purchasing on-premise cycles in Purdue's Community Cluster program \cite{McCartney2014}. This program provides subsidized computing and storage to faculty for only the capital cost of compute nodes. The expenses of the facility, staff, interconnect fabric, networking, etc. were subsidized with general funds through the Research Computing department. Although REED was also substantially subsidized, researchers were directly charged at an annual rate plus they covered project specific runtime expenses generated in the cloud.

In addition to different cost models, REED and a Community Cluster operated very differently. The Community Clusters provide a primarily Linux-based experience pre-loaded with a software stack the computing center has curated for over twenty years specifically for Purdue researchers. The REED environment required every software package to be individually requested, vetted by security, documented, and installed by Purdue staff. The Community Clusters are also extremely inviting for data storage/sharing, group management, and provide large default storage quotas (starting at 100TB per user on our oldest clusters). The REED environment required two people to approve file transfers through a cumbersome process and every byte of storage had to be watched carefully to avoid extreme cost fluctuations.  Not leveraging the university's investment in community clusters meant that solutions for CUI miss out on economies of scale, automation, self-service tools \cite{Colby2014}, and a consistent user experience with other computing resources.

As staff considered these key differences and others, it became clear the next generation resource supporting CUI research had to be different than REED while at the same time providing the security expected by the NIST SP 800-171 requirements.

\subsection{Options Explored}
After inviting feedback, this cross-section group of researchers and various IT support groups researched other front runners within the CUI space. San Diego Supercomputer Center offers Sherlock\cite{sherlock} that handles the security controls as a service. Notre Dame \cite{ndhybrid} is running a hybrid local and AWS GovCloud system which pulls the most expensive resources out of the cloud and back to campus while leaving central services. University of Florida \cite{FLresVault} also stood out as leaders with their ResVault solution run with software, tiCrypt. All three of these institution's solutions are formidable, but moving closer to our Community Cluster program to leverage our existing expertise, cost model, and processes looked to best streamline and reduce costs.

\subsection{Community Cluster Adaptations}
The first key change to the Community Cluster design was to provide isolation from non-US persons working at the University. While Purdue University values diversity and our workforce is highly inclusive when it comes to nationality, this provided a unique challenge in regards to hosting a CUI environment on campus. This was accomplished by documenting controls and revising processes for one of our data centers.  This data center is a self-contained system in the form of a shipping container. It features multi-factor entry systems enabled with access list limited to US persons. But the isolation did not stop here, it continued through the deployment of VPN firewalls, mirrored software repositories, replicated configuration management, and all the subsystems that make up an HPC cluster.

Next, we had to eliminate some of the features normally offered with the Community Clusters. These included services like Open OnDemand, Globus file transfer service, MySQL, and Hadoop (through Magpie) that started network services for users.  While data maintained on one Community Cluster is reachable from the others, due to cross-mounted file systems, that enables interoperability among research groups, data on Weber is entirely inaccessible to the other Community Clusters. 

Finally, we evaluated the remaining NIST requirements and implemented a process for file transfers, multi-factor authentication, logging, vulnerability detection, restricted copy/paste, etc.

A benefit to reusing the architectural design of the Community Clusters for Weber is the familiarity for staff plus the ability to leverage existing cluster deployment techniques. In the end, this all drove the cost to faculty for Weber down by half!

\subsection{Description of the Weber Cluster}
The Weber cluster consists of several key systems: an independent SFTP server for file ingress and egress, login nodes (or front-ends), computational nodes that enable distributed computing, and a number of infrastructure machines to support the cluster as a whole.  These components will be described in turn below and are depicted in Figure 1.  

\subsubsection{VPN Access for Users and System Administrators}
Access to Weber is controlled via two VPNs.  System administrators have access to a restricted set of administrative components to maintain the operating system, manage user permissions, install and maintain applications, etc.  The other VPN is accessible to users which prevent access to the above administrative servers, while allowing access to their needed services.  VPN is accessible only to U.S. persons who are part of a project's Technology Control Plan (TCP) coming from the geofenced U.S.
\begin{figure}[H]
 \includegraphics[width=.53\textwidth]{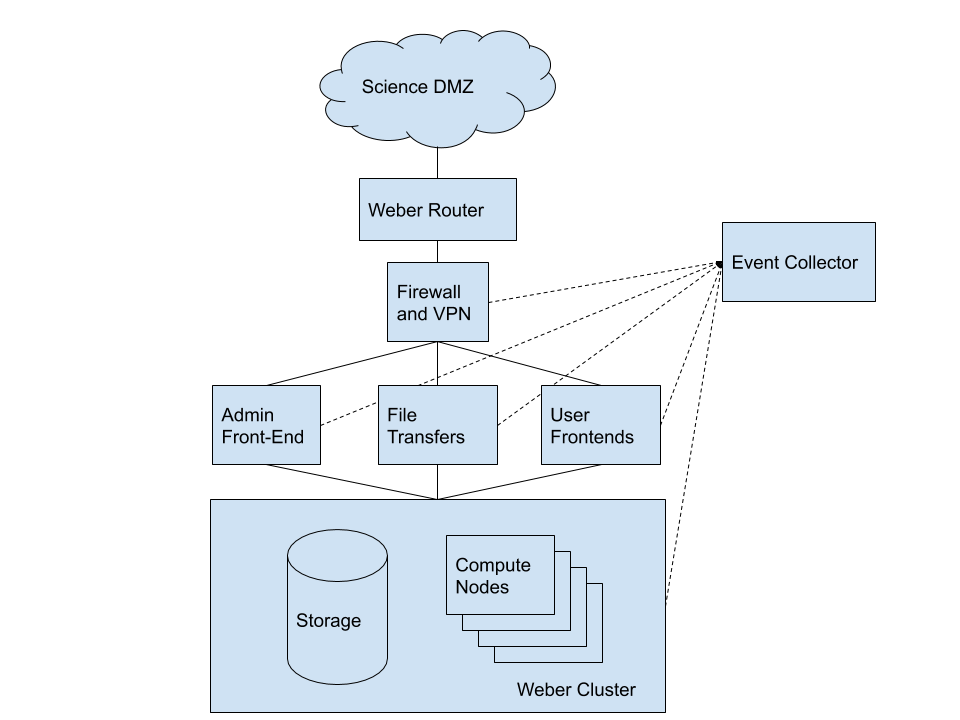}
  \caption{Weber Cluster Infrastructure Diagram.  Data originates from PIs, users, and project sponsors (Science DMZ).  The Event Collector tracks file access for later security auditing.}
%  \Description{Student competitor preparing for a Student Cluster Competition using the LittleFE environment.}
  %\label{fig:teaser}
\end{figure}

Access to the user VPN permits access to Weber's SFTP server, front-ends, and nodes.  The Weber SFTP subsystem is entirely independent from the Weber front-ends and nodes.  Once on the VPN, principal investigators (PIs) or select authorized users may upload or download files, into specially marked directories, i.e. ``inbox-outside'' and ``outbox-outside,'' respectively.  Starting with the ingress of files, a user uploads files into their own personal directory in the ``inbox-outside'' folder.  Once they have completed uploading files that they wish to compute on or store within Weber, they upload a specifically named file to start the transfer.  Every five minutes, the system searches for files within the ``inbox-outside'' folder that are accompanied by the transfer-flag file, then proceeds to scan them for malware.  Clean files are then made available to the user \emph{inside} Weber in their ``inbox-inside'' directory.  This directory is only accessible via the front-ends.

A mirrored version of this process enables the egress of data from Weber.  From the Weber front-ends, users copy or move data to the ``outbox-inside'' directory.  A transfer flag file is used to authorize the egress of data, which are then scanned for malware and made available in the ``outbox-outside'' directory on the Weber SFTP server. Figure 2 depicts the data flow within Weber.

Files may also be made available to project sponsors via outbound HTTPs drop sites.  These sites are white-listed by the firewall, after being reviewed by Export Control and security offices.

\begin{figure}[H]
 \includegraphics[width=.53\textwidth]{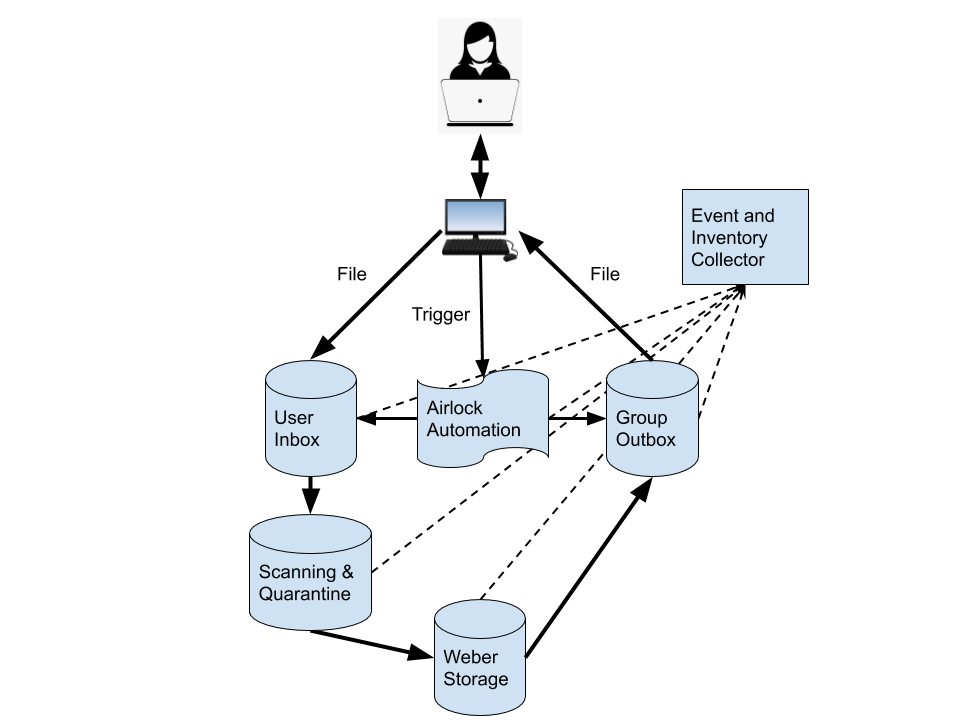}
  \caption{Weber Data Flow.  From left to right, users experience three major ways to interact with data: through the ThinLinc remote desktop environment, via data ingress, and via data egress.  All transactions are monitored by the Event and Inventory Collector.}
%  \Description{Student competitor preparing for a Student Cluster Competition using the LittleFE environment.}
  %\label{fig:teaser}
\end{figure}
\subsubsection{Local Hardware}
The two major subsets of hardware components on Weber are the front-end and compute nodes, with properties as summarized in Table 1.  The two front-end or login nodes are generally where most of the work in the CUI environment is performed.  These front-ends contain 16 2.10 GHz Sky Lake cores with 192 GB RAM per front-end.  Many users simply need a system that they may use to aggregate and produce data analysis results such as data tables and charts, and place them in documents for export from Weber to sponsors. These front-ends are also leveraged for data exploration and software prototyping.

Supporting these front-ends is a small cluster of 4 compute nodes, which feature 20 2.60 GHz Haswell cores and 64 GB of RAM per node.  This array enables longer-duration, more demanding computations to be offloaded from the front-ends to minimize the impact to other users.  Job scheduling on these nodes is performed by SLURM.  The most common code used both interactively and in batch is MATLAB and the Parallel Computing Toolbox.

\begin{center}
	\begin{table}[H]
		\caption{Key hardware component properties of the Weber system.}
		\begin{tabular}{|l|l|l|}
		%\tablecaption{Weber Hardware Properties}
			\multicolumn{1}{|c|}{\bf Properties} & \multicolumn{1}{c|}{\bf Front-end Nodes} & \multicolumn{1}{c|}{\bf Compute Nodes} \\
			\hline
			Number of Nodes             & 2               & 4             \\
			Cores per Node              & 16              & 20            \\
			Processor Microarchitecture & Sky Lake        & Haswell       \\
			Clock Speed (GHz)           & 2.10            & 2.60          \\
			RAM per Node (GB)           & 192             & 64            \\
		\end{tabular}
	\end{table}
\end{center}

\section{System Components and User Outcomes}

\subsection{Data Portability}
A common concern among users of the REED system was that two users were needed to approve the egress of files from the system. This was due to our understanding of the conservative technical implementation control \S 3.1.4 "Separate the duties of individuals to reduce the risk of malevolent activity without collusion."\cite{nisthandbook162}  As many projects continued to revise reports and data products up until the deadlines set by sponsors, there were many cases where users were unable to locate a second person to approve the egress of materials.  The SFTP server mitigates these concerns by permitting a single user with elevated access to remove files from Weber.  Security of the system is not compromised, however, as access logs and additional training are done to track user actions.

\subsection{VPN and Split Tunnelling}
One of the biggest contentions the users face when using a NIST SP 800-171 aligned resource is the forced prevention of Split Tunneling while connected to the VPN (NIST 800-171, 3.13.7: Prevent remote devices from simultaneously establishing non-remote connections with the
information system and communicating via some other connection to resources in external networks. \cite{nist3137}).  While this is understandably difficult to work with, overall, users have recognized its importance into maintaining the data security of a CUI based resource.

\subsection{Simultaneous, Multi-user Front-ends}
The relatively powerful front-ends that form the core computational capability of Weber offer several advantages.  First, several users, whether from the same group of various groups, may simultaneously use a single front-end node.  Their usage of memory and cores is controlled via cgroups. This prevents a single user from abusing the system.  Since several users are simultaneously sharing a resource, this yields a second advantage, cost savings.  A third advantage is that systems administrators spend less time maintaining shared front-end systems and do not need to individually tailor them for particular projects.

\subsection{High Performance Computing}
Within the AWS REED environment, high performance computing was provided by the deployment of virtual machine (VM) images which could then be replicated across a user-defined number of instances.  These instances varied in cost and capability, with instances that provided more system memory and higher core counts being available at higher cost.

However, in our experience only one of our research groups leveraged this distributed computing capability of AWS REED and it was an extremely costly endeavor with their monthly bill averaging near \$4500.

Today, HPC capabilities on Weber are critical for future growth, with multiple pending contracts in the pipeline for faculty who are heavy users of HPC. These scientists use HPC resources for tightly coupled CFD simulations.

\subsection{Graphics Processing Units (GPUs)}
Similar to the HPC instances described above, AWS REED made available instances backed by GPU hardware at additional cost. While we explored use cases for Weber requesting GPUs, within REED it was always too cost prohibitive and never made it past estimates of monthly expenses. Being able to add GPU nodes to Weber as demand increases may be incredibly beneficial to future researchers.

\subsection{Unified Desktop Experience}
The mechanism that customers use to remotely access and manipulate their data on Weber is via the ThinLinc client installed on university-managed computers.  The client provides a graphical desktop interface to the Weber Linux environment, with menus, toolbars, icons to access common applications, and a terminal.  This client natively supports X Window System (X11) for graphical interfaces and enables sessions and processes to be left running on Weber front-ends, independent of the state of the user's machine. Both ThinLinc and X Windows interfaces also serve to assist new non-HPC users transitioning into Weber after being subject to the NIST SP 800-171 requirements. Applications and menus are arranged in the same manner on Weber as on our non-export controlled Community Clusters, which provides users with a single, familiar desktop experience across all clusters.

\subsection{Long-term Storage}
Once an export-controlled project is completed, it transitions into ``long-term storage.''  Prior to the construction of Weber, CUI data was stored in three ways: inside the REED environment, in encrypted archives placed into campus storage, and on encrypted hard drives locked within storage cabinets.  Data stored within the REED environment represents a continuous, metered cost with users paying AWS for storage per month. This model was particularly prohibitive to projects that had lost their current funding or were seeking future funded opportunities. Alternatively, archives were stored on the campus enterprise storage, through an encryption software Vormetric.\cite{vormetric}  However, Vormetric is an enterprise software that requires annual service fees and regular maintenance for REED's very small number of impacted projects. The third alternative, storing encrypted drives in locked cabinets, resulted in inefficiencies in security auditing and usage.  Security personnel were required to annually access locked drives to verify their location and contents.  Users found that the drives needed to be located, decrypted, and installed onto an air-gapped, standalone system for use to acquire future funding.

Our on-prem, hardware-based solution obviates many of these concerns. On Weber, long term data storage is built into the system.  Once projects are finished, the Export Control Office (ECO) simply requests file permissions be modified to remove access to the project directories and files for everyone except admin personnel.  The project directories remain within Weber and are flagged for additional logging and alerts of access or modification.

When PIs require prior project results to create a new proposal, the ECO can restore access with a simple request. Another advantage of this method is that within the REED environment, the creation of project proposals required the initialization and usage of instances which cost money. Within the Weber system, new proposals can be authored at little additional cost and with minimal intervention of staff.

\subsection{Reduced Cost Variability}
REED researchers could be surprised by their monthly bill for AWS since services were billed by instance, CPU-hours used, and data storage costs. Weber billing involves a flat “pay to play” annual rate that absorbs access, data storage, and computation costs. This makes the costs easier to budget, reduces month-to-month variability, and reduces anxiety about how computations may affect budgeted computing resources.

\subsection{User Accounts and Access}

After aggregating support tickets and emails that requested assistance with the prior REED environment, we were able to categorize their contents to search for common themes. A key discovery was that approximately 24\% of email threads to REED support staff and 12\% of tickets requested password resets. REED requires users to reset their passwords every 60 days, but a common error was that users would neglect to change their passwords in a timely manner and become locked out of their account. By leveraging two-factor authentication on the VPN and aligning to the centrally managed University credentials and other Purdue information systems, our migration to Weber removes this common source of user frustration.
After a project is awarded, and a Technology Control Plan (TCP) is developed from the ECO, they are ready to get access to Weber. Leveraging the user portal, each project gets a group that membership is entirely managed by ECO staff. The benefit to this separation in duties of handling group membership of the project. ECO staff are the closest to the information when a background check has been completed or as project members leave the project. ECO staff are able to immediately address and TCP documentation. By streamlining surrounding CUI business processes makes getting access to Weber more efficient process. 

\subsection{Windows Virtualization}
While the majority of research is done using Linux based applications, there are some cases where a Windows application is required (e.g., SolidWorks) or simply preferred (e.g., Microsoft Office versus LibreOffice). Such applications are implemented using a VM managed by QEMU using copy on write (COW) mode.  The VM emulates a Windows 10 environment supported by a default of 8 GB RAM and 2 CPU cores, although the supporting hardware is configurable by the user.  We have had this offering on our Community Clusters for some time, however for Weber we needed to re-design the implementation with NIST SP 800-171 requirements in mind. Users no longer have full administrative access to their Windows VM to prevent unauthorized software installation and service configuration. Additionally, a user VM is stateless, it is created from a COW of the "golden-image" and has access to the users home directory. Access to the Weber storage is provided inside the VM through QEMU's built-in file sharing using a private instance of Samba. VM's are private to individual users and the Samba instance is only available on virtual networking isolated to the VM. When the user logs off, or otherwise closes their VM, it is deleted. This assures that CUI data does not get stored in a non-monitored file system.

The most common software requests for Windows Virtualization on Weber are for Microsoft Office and MATLAB. If users require more computational power, the VMs can be launched on compute nodes through SLURM.

\subsection{File Permission Isolation}
File permissions management is a core control in Weber. For cost reasons, the centralized storage is shared between multiple projects who must not access each other's data. To ensure proper isolation, the top level directories in all spaces are owned by the root user and a group in which authorized users reside. The authority to administer group members resides solely with the ECO.  For instance, members in Lab A share data in a directory located at /depot/laba/data. The directories /depot and /depot/laba are not writable by anyone but root and the permissions on /depot/laba are mode 750. Only the inner directory /depot/laba/data is where users may place files. For all home, scratch, group, and archive spaces these restricted directory permissions are set at creation time and monitored frequently. Any abnormal changes to file permissions are logged and reviewed.

\section{Conclusion and Future Work}
Although Weber represents an advance for our institution over the previous REED system, opportunities remain to provide enhanced computation and data storage services to our user base.  One request that has repeatedly cropped up since the deployment of Weber is to enhance the user experience for the Windows Virtual Machine images. This request consists of two components: reducing the initialization times of the images and enhancing the responsiveness of the VM.  First, as the VM images are $\sim$70 GB in size, they can require minutes to copy, load into memory, and initialize for usability.  Second, users have reported a sluggish response from applications within the VM, including Microsoft Office.

Although Weber was created with clients that work with export-controlled data in mind, many interested users have requested that we deploy similar computation and data storage services to assist our life sciences user base, including professions such as bioinformatics, nursing, psychology, and veterinary medicine.  Another opportunity lies in developing a cluster that could facilitate the analysis of even more restricted data, such as those that may only be accessed by users with security clearances.

\begin{acks}
We thank our research and faculty partners including Daniel DeLaurentis, Kris Ezra, and Robert Campbell for their feedback on Research Computing resources that have helped us to improve the performance, reliability, and usability of the AWS GovCloud and Weber systems. MR would like to thank Saurabh Bagchi (Purdue University) and Thomas Dover (Butler County Community College) for helpful comments that have improved this work.  Some of this effort has been supported by NSF Advanced Cyberinfrastructure Award \#1840043.
\end{acks}

\bibliographystyle{ACM-Reference-Format}
\bibliography{sample-base}

%\appendix

\end{document}